\documentclass[]{aastex631}

\definecolor{darkgreen}{rgb}{0,0.50,0}
\shorttitle{Fullerenes in AhS}
\shortauthors{Sabbah et al.}
\graphicspath{{./}{figures/}}
\begin{document}

\title{Detection of cosmic fullerenes in the Almahata Sitta meteorite: are they an interstellar heritage?
} 

\correspondingauthor{Christine Joblin, Hassan Sabbah}
\email{christine.joblin@irap.omp.eu, hassan.sabbah@irap.omp.eu}

\author[0000-0001-5722-4388]{Hassan Sabbah}
\affiliation{IRAP, Université Toulouse III - Paul Sabatier, CNRS, CNES  \\ 31028 Toulouse Cedex 4, France}

\author{Mickaël Carlos}
\affiliation{IRAP, Université Toulouse III - Paul Sabatier, CNRS, CNES  \\ 31028 Toulouse Cedex 4, France}

\author[0000-0003-4735-225X]{Peter Jenniskens}
\affiliation{SETI Institute \\ Mountain View, California 94043, USA}

\author{Muawia H. Shaddad}
\affiliation{University of Khartoum \\ Khartoum 11115, Sudan}

\author[0000-0001-7408-3089]{Jean Duprat}
\affiliation{IMPMC, CNRS-MNHN-Sorbonne Université \\
57 rue Cuvier, 75005 Paris, France}

\author[0000-0002-9820-3329]{Cyrena A. Goodrich}
\affiliation{Lunar and Planetary Institute  \\ USRA, Houston, TX 77058, USA}

\author[0000-0003-1561-6118]{Christine Joblin}
\affiliation{IRAP, Université Toulouse III - Paul Sabatier, CNRS, CNES  \\ 31028 Toulouse Cedex 4, France}

\begin{abstract}
Buckminsterfullerene, C$_{60}$ is the largest molecule observed to date in interstellar and circumstellar environments. The mechanism of formation of this molecule is actively debated. Despite targeted searches in primitive carbonaceous chondrites, no unambiguous detection of C$_{60}$ in a meteorite has been reported to date. Here we report the first firm detection of fullerenes, from C$_{30}$ to at least C$_{100}$, in the Almahata Sitta (AhS) polymict ureilite meteorite. This detection was achieved using highly sensitive laser desorption laser ionization mass spectrometry. Fullerenes have been unambiguously detected in seven clasts of AhS ureilites. Molecular family analysis shows that fullerenes are from a different reservoir compared to the polycyclic aromatic hydrocarbons detected in the same samples. The fullerene family correlates best with carbon clusters, some of which may have been formed by the destruction of solid carbon phases by the impacting laser. We show that the detected fullerenes are not formed in this way. We suggest that fullerenes are an intrinsic component of a specific carbon phase that has yet to be identified. The non-detection of fullerenes in the Murchison and Allende bulk samples, while using the same experimental conditions, suggests that this phase is absent or less abundant in these primitive chondrites. The former case would support the formation of fullerenes by shock wave processing of carbonaceous phases in the ureilite parent body. However, there are no experimental data to support this scenario. This leaves open the possibility that fullerenes are an interstellar heritage and a messenger of interstellar processes.

\end{abstract}



\section{Introduction} 
\label{sec:intro}

Kroto and co-workers discovered the C$_{60}$ buckminsterfullerene in the laboratory during the synthesis of carbon clusters (C clusters) of interest for carbon chemistry in evolved stars \citep{Kroto1985}. The spectral signatures of C$_{60}$ and its cation C$_{60}^{+}$ have subsequently been identified in the diffuse interstellar medium \citep{Campbell2015, Berne2017} as well as in a variety of circumstellar and interstellar environments, which include planetary nebulae \citep{Cami2010, Garcia-Hernandez2012,Otsuka2014} and photodissociation regions \citep{Sellgren2010, Castellanos2014}. Different scenarios have been proposed to account for C$_{60}$ formation in these environments, including processing of grains by shocks or high-energy ions \citep{Scott1997, Otsuka2014, Bernal2019}, and UV photo-processing of large polycyclic aromatic hydrocarbons (PAHs) \citep{Berne2012, Zhen2014, Berne2015, Berne2016}. The most energetic conditions are expected to favor the most stable species, buckminsterfullerene C$_{60}$.  

Recent progress has been made in the laboratory to mimic the gas phase chemistry in the environment of evolved stars using the dedicated Stardust machine \citep{Martinez2020Prevalence}. However, even with these improved methods, no fullerenes could be formed from reactivity of a carbon vapor (C/C$_2$) with H$_2$ and C$_2$H$_2$ \citep{Martinez2020Prevalence, Santoro2020chemistry}. This is likely due do the relatively low gas temperature during aggregation in the Stardust machine, which is estimated to be $<$1,000 K. Indeed, it has been shown in the laboratory that the formation of fullerenes requires temperatures above 3500~K in a gas-phase condensation experiment \citep{Jaeger2009}. The “shrinking hot giant” pathway has been demonstrated by quantum chemical molecular dynamics simulations of the dynamics of carbon vapour \citep{Irle2006}. In this scenario, the formation of fullerenes results from the assembly of hot polyyne chains followed by shrinking of the vibrationally excited cages towards the sizes of C$_{60}$ and C$_{70}$. In astrophysical environments, gas-phase formation of C clusters, fullerenes, and graphite grains has been modeled by Clayton et al. in the  C + O cores of core-collapse supernovae  \citep{Clayton2001, Clayton2018}, although the authors limit their model to temperatures below 2000~K.  \cite{Cherchneff2000Dust} have proposed the formation of the same species in the pre-supernova stage of Wolf Rayet (WR) stars. These stars have different phases including a carbon-rich phase (WC stage) that can produce very large amounts of carbon dust \citep{Crowther2003Dust}. The extreme conditions encountered in these environments combined with a very rich medium in carbon could thus be favorable to the formation of C clusters and fullerenes  \citep{Cherchneff2000Dust}. However until now no link has been established between this carbon chemistry modeled in massive stars and the observed C$_{60}$.

Because of their stability under UV irradiation, C$_{60}$ in particular, and fullerenes in general, are good candidates to survive the journey from stars and the interstellar medium to our Solar System (SS) and would be naturally incorporated into SS solid bodies.
Becker et al. searched for fullerenes in two primitive carbonaceous chondrites, Allende and Murchison \citep{Becker1994, Becker1997, Becker1999Higher, Becker2000}. The authors combined chemical extraction to increase the concentration of fullerenes with one-step laser desorption ionization (LDI) mass spectrometry. They reported first the detection of C$_{60}$ and C$_{70}$ \citep{Becker1994} and then that of larger fullerenes 
However, the level of detection was found to be highly variable between samples of the same meteorite \citep{Becker1997}. Moreover, their detection could not be confirmed by other groups \citep{Buseck2002}. \cite{Hammond2008Identifying} demonstrated that the detected fullerenes were not intrinsic to the samples but generated by the one-step LDI process used to analyze the samples. A similar conclusion was reached in a recent study of insoluble organic matter in the Paris meteorite by \cite{Danger2020}.
\cite{Buseck2002} concluded that the detection of fullerenes in meteorites is a difficult task that must combine careful molecular extraction, analysis close to the limits of sensitivity, and exquisite care to avoid contamination at very low concentration levels.

In the experiments mentioned above, a single laser was used to perform both desorption and ionization (one-step LDI). In the two-step laser desorption laser ionization mass spectrometry (L2MS) technique, desorption and ionization are separated in both time and space using two different lasers. Compared to one-step LDI, this technique provides better control of the laser desorption fluence and increases sensitivity by one to two orders of magnitude. The desorption laser can be used at a lower fluence, thereby limiting chemical interactions in the desorbed plume, thus circumventing the ambiguities of LDI analysis raised by previous measurements by Becker et al. and also  described in \cite{Danger2020}. The new experimental setup AROMA (The Aromatic Research of Organics with Molecular Analyzer) combines L2MS with ion trapping. We have demonstrated the ability of AROMA to detect PAHs and fullerenes (C$_{60}$) with high sensitivity (up to 100 femtograms per laser desorption shot) and with almost no fragmentation \citep{Sabbah2017Identification}. The PAH distribution we obtained from the Murchison meteorite, is consistent with previous studies \citep{Callahan2008} but with improved mass resolution and sensitivity, particularly for species with m/z greater than 200. Combining the double equivalent (DBE) method and collision-induced dissociation experiments, we identified the dominant peak from the Murchison analysis at m/z=202.07 as pyrene. Series of methylated species were also identified. In this earlier study, no C clusters or fullerenes were detected.

L2MS analysis was previously applied to nine samples of the Almhata Sitta (AhS) polymict ureilite meteorite \citep{Sabbah2010Polycyclic}. The authors confirmed the presence of PAHs in AhS and their dispersion among a variety of clast types, and concluded that this dispersion results from impacts on the fragmented ureilite parent body (UPB).
Interestingly, some of the recorded m/z peaks indicated the possible presence of C clusters or aromatics (not the usual standard PAHs), but further assignment was not possible due to lack of sufficient mass resolution. These observations motivated us to undertake a detailed systematic analysis of 13 AhS samples using the new AROMA L2MS setup to explore chemical diversity across different samples of different lithologies.

The AhS meteorite was recovered after the impact on Earth of the asteroid 2008 TC$_3$. This exceptional event was observed in space and followed until the final impact over the Nubian desert in October 2008 \citep{Jenniskens2009impact}. The collected samples were found to be a polymict ureilite. This is the first witnessed polymict ureilite fall collected shortly after the fall. The samples were recovered from a  well isolated environment with minimal terrestrial contamination.
They consist primarily of ureilites of various types, but also include various chondrites, including enstatite chondrites, ordinary chondrites, and carbonaceous chondrites (CC). The ureilites originated from the UPB, a carbon-rich planetesimal that formed in the inner SS, based on its place in the SS nucleosynthetic isotope dichotomy \citep{Warren2011}, at $<$1 Ma after the earliest solids, i.e., calcium–aluminium-rich inclusions (CAI), formed \citep{Wilson2008Thermal}. The UPB experienced rapid heating and partial differentiation, but was disrupted by a major impact at ~5-5.4 Ma after the CAI, before complete cooling \citep{Downes2008Evidence, Goodrich2015Origin, Goodrich2004Ureilitic,Herrin2010Thermal}. Subsets of fragments produced by the disruption reassembled by self-gravity \citep{Michel2015Selective, Michel2015Collisional}, forming daughter bodies from which the known ureilites derive \citep{Goodrich2015Origin}. These bodies subsequently developed complex regoliths including foreign remnants of impacting chondritic meteorites, some of which are preserved as the chondritic stones from AhS. Polymict ureilites, including AhS, are samples of these regoliths.

In this article, we report the detection of fullerenes in clasts of the polymict ureilite AhS, which is the first strong evidence for the presence of such carbonaceous species in a meteorite. The experimental setup and methodology are described in Section~\ref{sec:experimental}. Section~\ref{sec:Results} demonstrates the detection of fullerenes for 7 of the 13 samples of the AhS meteorite. Using the same experimental conditions, we were unable to detect fullerenes in the Murchison and Allende CC bulk samples. In section~\ref{sec:discussion}, we propose two scenarios to explain the presence of fullerenes in the AhS meteorite, an origin within the solar system (i.e. from shocks to the surface of the UPB) or an interstellar heritage.
We conclude in Section~\ref{sec:conclusion}.

\section{Experimental techniques} \label{sec:experimental}

\subsection{Sample preparation} \label{subsec:sample}

For this study 13 clasts from the AhS meteorite were used to track chemical diversity among them. Seven of them (AhS \#04, \#22, \#24, \#27, \#28, \#38, and \#48) are ureilites with a high concentration of aggregates of carbonaceous material (up to 500 microns in diameter) composed of graphite and minor diamonds \citep{Jenniskens2009impact,Zolensky2010}. The remaining six samples are enstatite and ordinary chondrites, which are classified as follows: AhS \#41 (EL6), \#58 (H4-5), \#1001 (Metal+Sulfide), \#1002 (EL4-5), \#1054 (LL3-4), and \#2012 (EH4-5). In addition, a bulk sample from the interior of the Allende CC was used.  
For each analysis fresh fragments of few mg were crushed using a mortar and pestle. The powder was then attached to a 10\,mm disc of stainless steel with conductive copper tape. The disk holding the powder was then mounted on the sample holder to be inserted inside the instrument via an automated vacuum interlock system. After introduction, the sample was positioned and moved in two axis using a motorized XY-manipulator with a minimum step of 100\,$\mu$m. It was demonstrated that the copper tape did not produce a background signal that could interfere with the analysis.

\subsection{The AROMA setup} \label{subsec:AROMA}

AROMA, the Astrochemistry Research of Organics with Molecular Analyzer, is a unique experimental setup developed to study, with micro-scale resolution, the content in carbonaceous molecules of cosmic dust analogues and meteoritic samples \citep{Sabbah2017Identification}. Mass spectrometry data and chemical analysis tools for all studied samples available to the public in the AROMA database at http://aroma.irap.omp.eu. 

\begin{figure*}[ht!]
\plotone{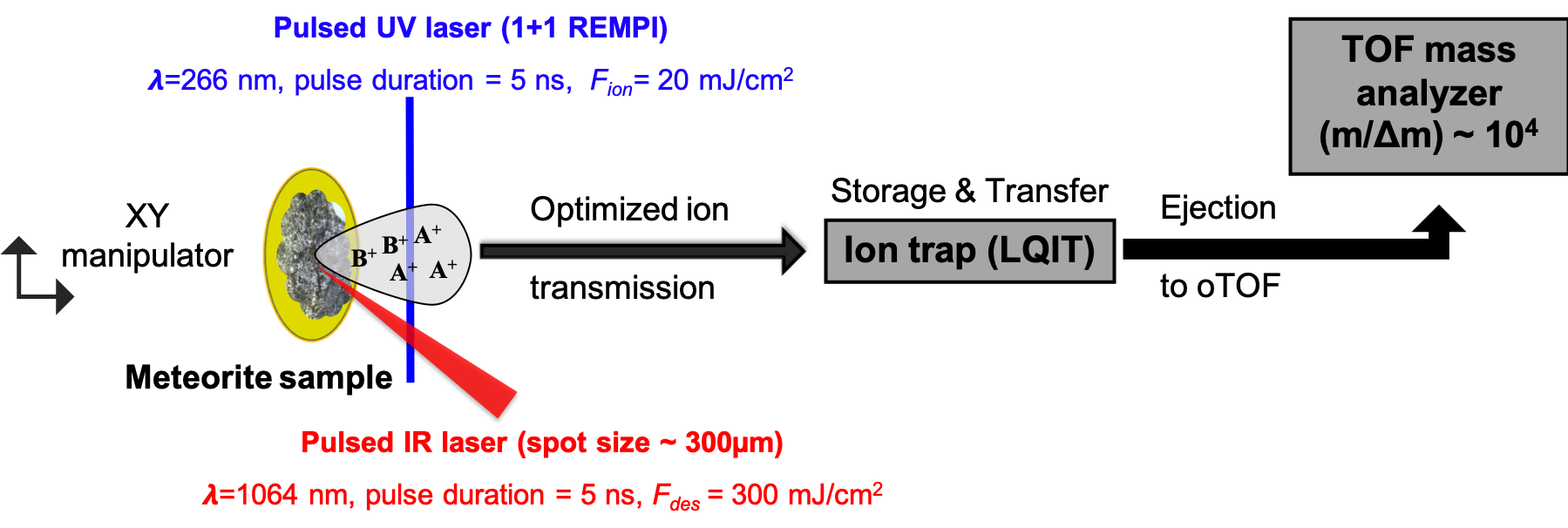}
\caption{Schematic diagram highlighting the two-step LDI scheme and the main components of the AROMA L2MS setup.
\label{fig:aroma}}
\end{figure*}

AROMA consists of a microprobe laser desorption ionization source and a segmented linear quadrupole ion trap (LQIT) connected to an orthogonal time of flight (oTOF) mass spectrometer, as shown in Figure~\ref{fig:aroma}. Ions are produced at a low background pressure (10$^{-6}$\,mbar) by performing a one- or two-step LDI.  The laser-generated ions are collimated by a set of lenses at high DC voltage and are thermalized in a radio-frequency (RF) octapole, which maximizes ion transmission and reduces fragmentation that occurs in a typical LDI source. The ions are stored in the LQIT and processed if desired. A set of RF-DC optics is used to transfer the ions to high vacuum and finally they are monitored on an oTOF mass analyzer equipped with a two-stage reflectron and a fast microchannel plate (MCP) detector. The total mass spectrum of an experiment is the superposition of multiple scans recorded over a given m/z range. The amplitude and RF frequency applied to the LQIT electrodes are optimized to maximize the ion signal in this mass range. Each scan is the result of 50 laser shots. The interaction of the desorption laser with the sample is controlled by a mechanical shutter. The laser hits the sample once every two seconds. The sample is then moved so that the next laser shot hits a fresh spot. 

In order to perform LDI, AROMA uses a pulsed (5\,ns) infrared (IR) laser (Nd:YAG at 1064\,nm) focused on the sample with a spot size of 300\,$\mu$m to cause rapid and localized heating, promoting thermal desorption rather than decomposition. The typical IR laser desorption fluence used in this work is \emph{F}$_{des} = 300$\,mJ/cm$^{2}$ (60\,MW/cm$^{2}$ in irradiance). This low fluence is one to two orders of magnitude lower than the explosive vaporization threshold that leads to plasma formation during ablation processes \citep{Hoffman2014effect}. The interaction of the laser with the sample is able to efficiently produce ions from the metallic phases, which are present in minerals, salts and the matrix of natural samples \citep{Jayasekharan2013Elemental,Koumenis1995Quantitation,Shi2016Recent,Tulej2011miniature}. In this step, C clusters can also be generated from the decomposition of a pure carbonaceous phase \citep{Sedo2006}.
In the L2MS analysis, a pulsed (5\,ns) ultraviolet (UV) laser (fourth harmonic of an Nd:YAG at 266\,nm) perpendicularly intercepts the expanding plume. This leads to the selective ionization of species that can undergo (1+1) resonance-enhanced multiphoton ionization (REMPI), which is the case of fullerenes and aromatic species. The laser ionization fluence used in this work is \emph{F}$_{ion} =20$\,mJ/cm$^{2}$ (4\,MW/cm$^{2}$ in irradiance). 

\begin{figure*}[ht!]
\plotone{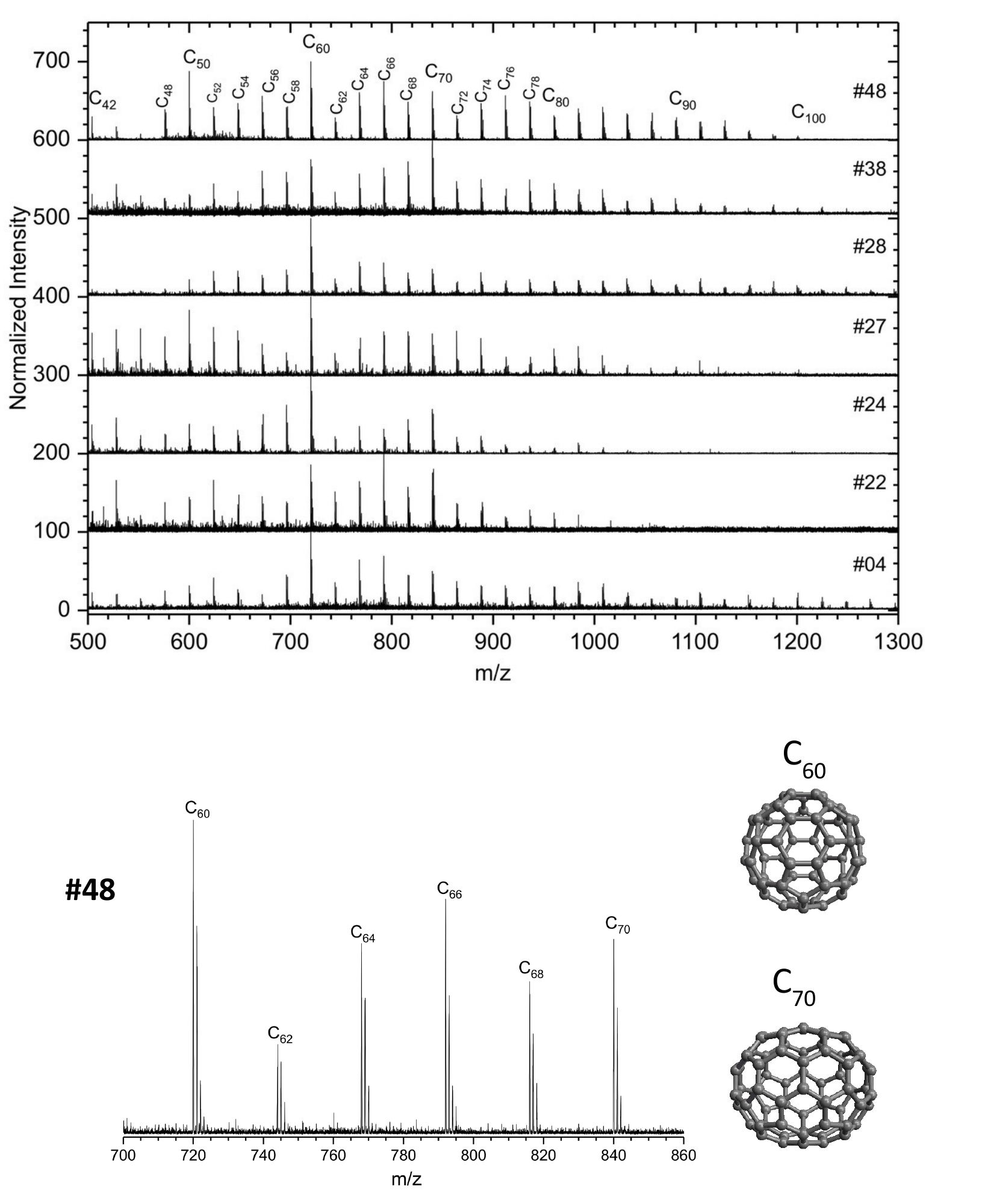}
\caption{Mass spectrum of ureilites (AhS \#04, \#22, \#24, \#27, \#28, \#38, and \#48) in the high m/z range of 500 to 1300. The observed peaks are attributed to fullerenes. In the lower left, a zoom on the C$_{60}$ and C$_{70}$ peaks shows the likely contribution of $^{13}$C isotopic species. The structures shown for C$_{60}$ and C$_{70}$ are from http://www.nanotube.msu.edu/fullerene/fullerene-isomers.html. 
\label{fig:fullerenes_AhS}}
\end{figure*}

\subsection{Element and carbonaceous molecular family analysis} \label{subsec:DBE}

For each detected m/z peak with a signal-to-noise ratio (S/N) greater than 10, a chemical formula is assigned using mMass software, an open source mass spectrometry tool \citep{Strohalm2010}.
The chemical formula assignment is performed with a typical accuracy of 0.01 between measured and calculated m/z values.
Peaks corresponding to various metallic elements (most intense peaks: Fe, K, Na, Ca, Al, Cr…) seen as atomic ions or small clusters (e.g., Fe$_2$) have been identified thanks to their mass defect and the high mass resolution provided by AROMA. 
This family is referred to hereafter as “metals” and reflects the different silicate phases found in AhS. As is well known in LDI mass spectrometry, signal from elements, in particular Na and K,
is ubiquitous in natural and standard laboratory samples.
For the molecular family analysis, we calculate in addition the double bond equivalent (DBE) \citep{Marshall2008Petroleomics} for each molecular formula, which is defined as:
\begin{equation}
DBE = C\# - H\#/2 + N\#/2 +1,
\end{equation}
with C\#, H\# and N\#, the number of C, H and N atoms in molecules (in our analysis we only include C and H because the mass resolution of the AROMA setup is not sufficient to disentangle the mass of N from that of CH$_2$). The DBE is representative of the unsaturation level of the molecules and thus corresponds to a direct measure of their aromaticity. It is equal to the number of rings plus double bonds involving carbon atoms (as each ring or double bond results in the loss of two hydrogen atoms). In the recorded mass spectra, only a few species containing oxygen, not exceeding C$_7$, could be firmly identified.
The DBE calculation allows us to sort the detected pure carbon and hydrocarbon ions into different molecular families \citep{Martinez2020Prevalence, Sabbah2020Molecular}. Pure carbon species are sorted into C clusters (C\# $<$ 30) and fullerenes (C\# $\geq$ 30). The transition between C clusters and fullerenes is somewhat arbitrary. \citet{vonHelden1993} have shown that fullerenes appear in the C\# range of 30 to 40 but this does not mean that all species in this range are actually fullerenes. Hydrocarbons are classified into three categories, HC clusters, PAHs, and aliphatics species. This classification is done using empirical factors and DBE limits established for hydrocarbon mixtures in complex natural organic matter \citep{Hsu2011Compositional, Koch2006From, Lobodin2012Compositional}. Hydrocarbons with 0.5 $\leq$ DBE/C\# $\leq$ 0.9 are considered to be PAHs. HC clusters and aliphatic species are located at DBE/C\# $>0.9$ and $<0.5$, respectively.

\section{Results}
\label{sec:Results}

\begin{figure}[ht!]
        \centering
        \includegraphics[width=0.7\linewidth]{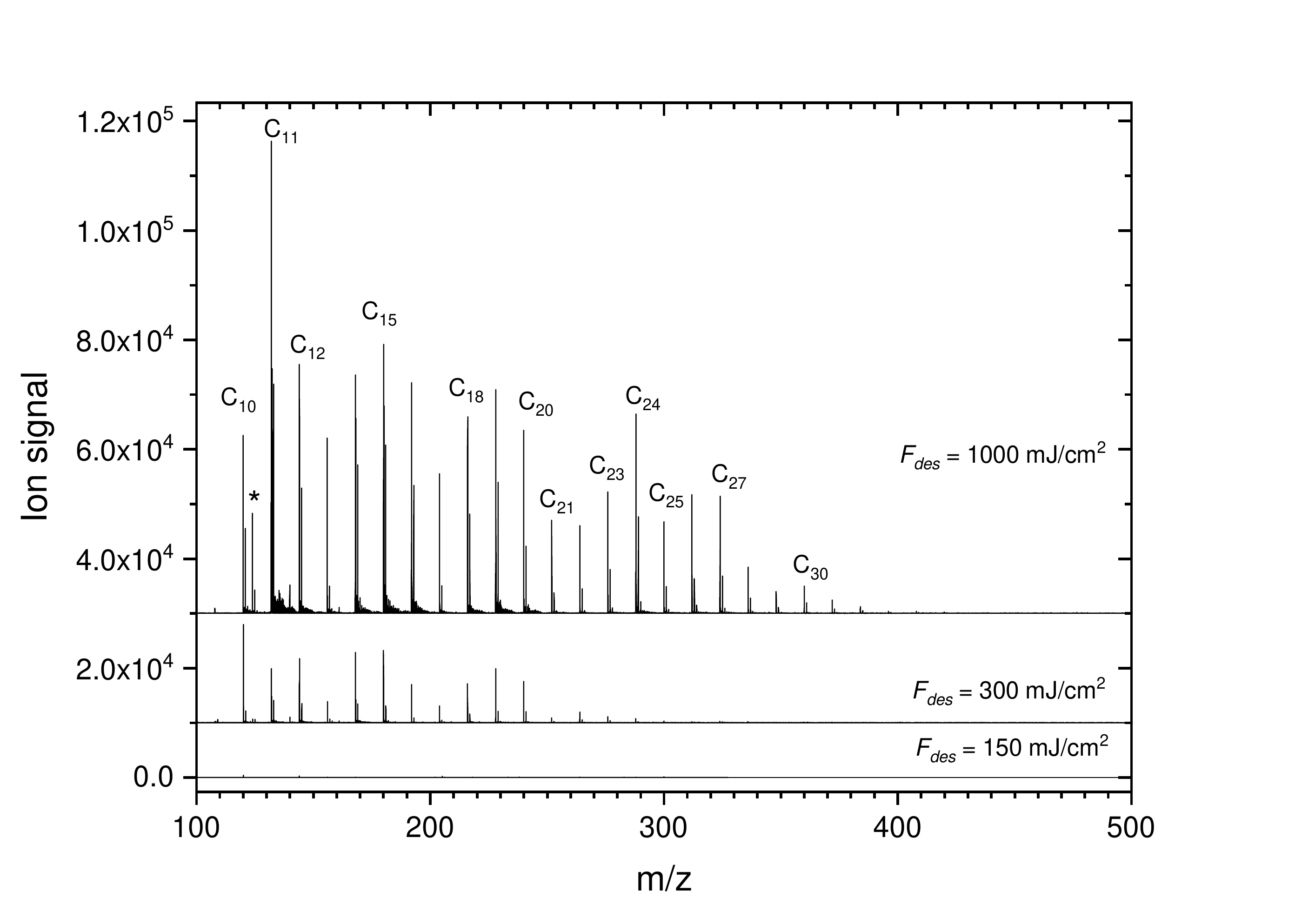}
\caption{Mass spectrum of HPOG recorded in the m/z range of 100-1000 at three different laser desorption fluence values, \emph{F}$_{des} = 150$, 300, and 1000\,mJ/cm$^{2}$ (one-step LDI scheme). The assigned peaks are labeled with the corresponding neutral species. * m/z = 123.98 and 124.98 are attributed to C$_{9}$O and C$_{9}$HO, respectively.
\label{fig:figHPOG}}
\end{figure}

\begin{figure}[hb!]
        \centering
        \includegraphics[width=1.0\linewidth]{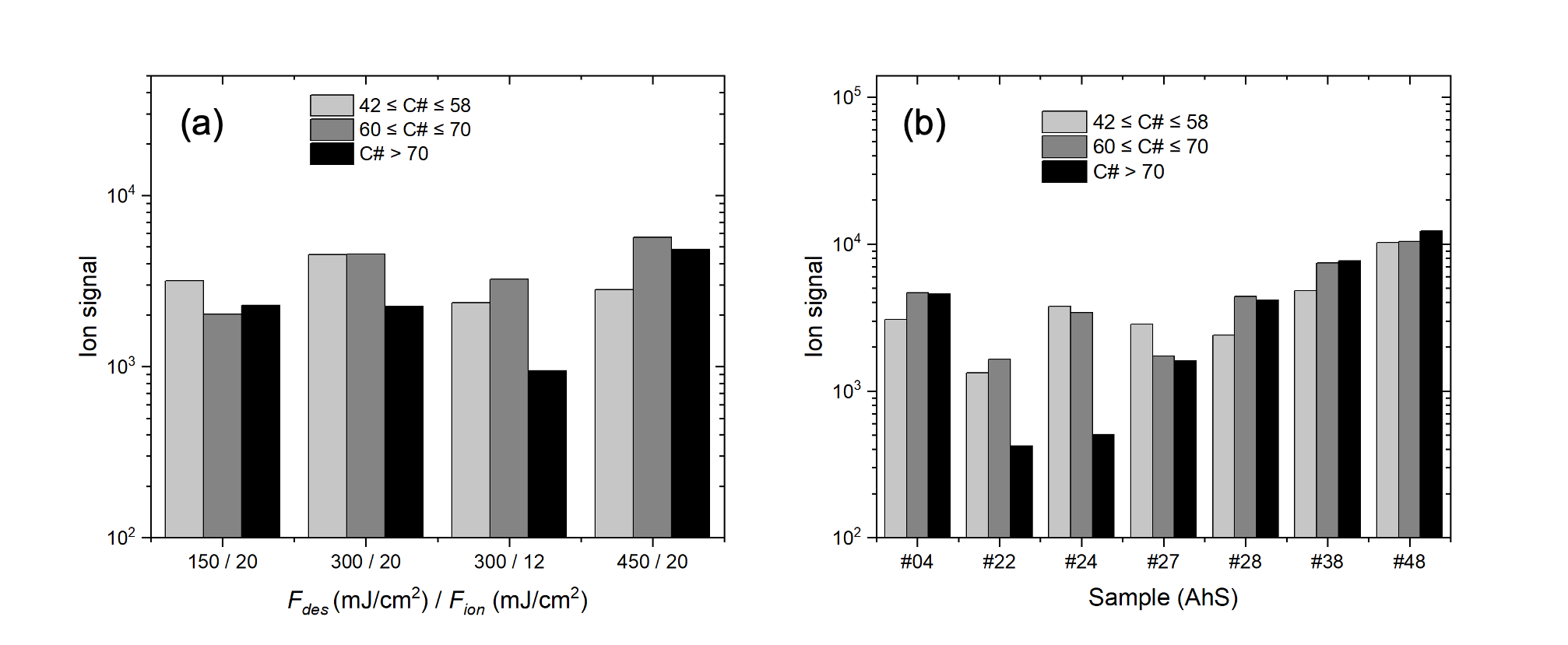}
\caption{Sum of peak intensities for fullerene species grouped in the three size ranges of medium ($42\leq C\# \leq58$), large ($60\leq C\# \leq70$), and very large ($C\#>70$). (a): Intensities for the AhS~\#04 sample as a function of laser fluences, \emph{F}$_{des}$ and \emph{F}$_{ion}$ (two-step LDI scheme). (b): Intensities for the seven ureilite samples.
\label{fig:fullerenes_cal}}
\end{figure}

\begin{figure*}[ht!]
\plotone{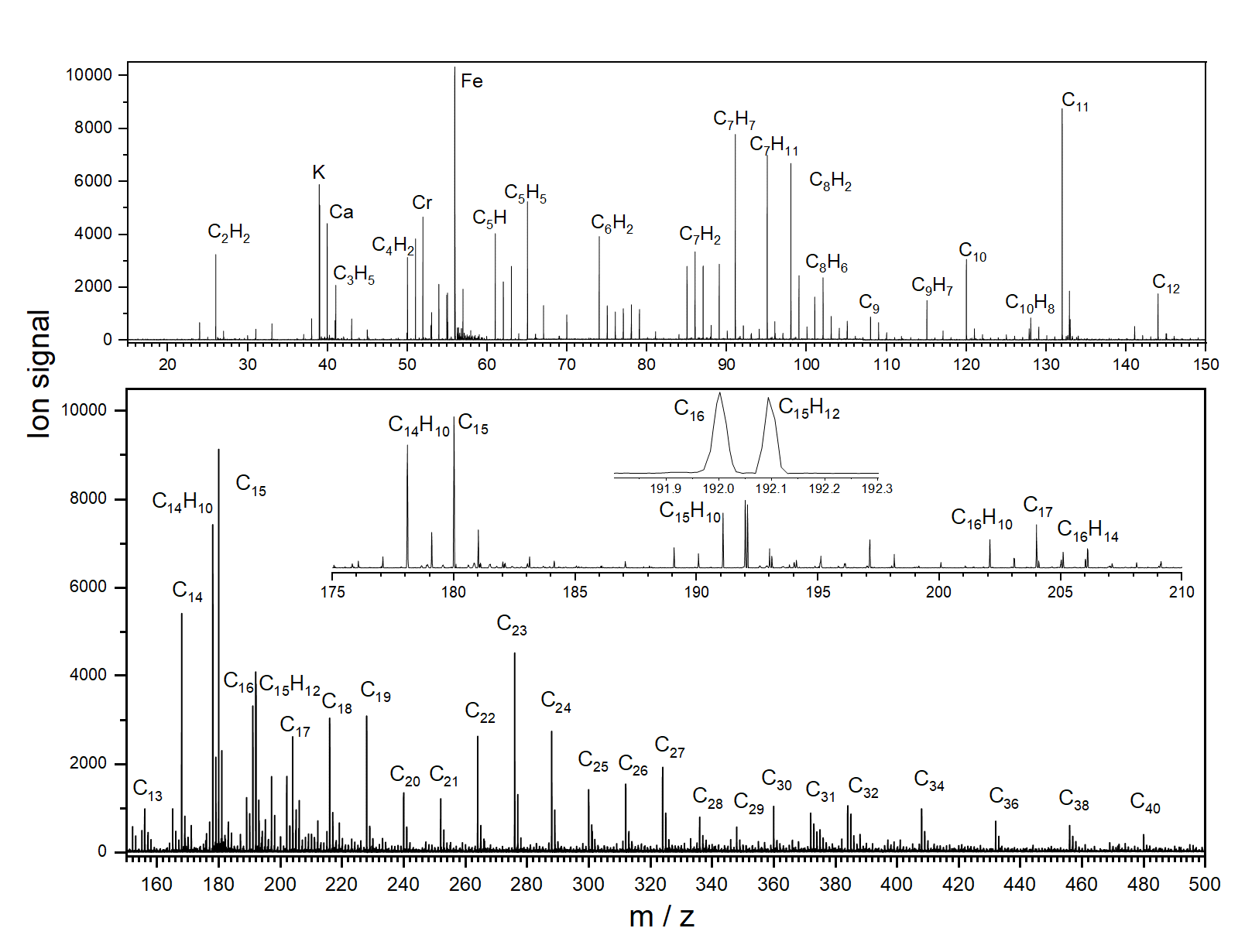}
\caption{Mass spectrum of AhS ureilite \#04 recorded in the low m/z range of [20 - 150] (upper panel) and [150 - 500] (lower panel), respectively. The first m/z range is dominated by metals and HC clusters, while the second m/z range is dominated by C clusters and PAHs. The assigned peaks are labeled with the corresponding neutral species. The resolving power of the instrument is illustrated in the case of the peaks at 192.00 and 192.09 corresponding to C$_{16}$ and C$_{15}$H$_{12}$. 
\label{fig:lowMS}}
\end{figure*}


\subsection{Detection of fullerenes in the AhS meteorite} 
\label{subsec:Fullerenes}

Figure~\ref{fig:fullerenes_AhS} gathers the data that were recorded in the range m/z=[500,1300] for the seven ureilites. We could not detect any ion signal in this range for the other clasts. Figure~\ref{fig:fullerenes_AhS} shows that species from C$_{42}$ to about C$_{100}$ (C$_{80}$ for \#22 and \#24) are observed with a mass spacing of 24 m/z (C$_{2}$), which is characteristic of fullerene series.
Figure~\ref{fig:figHPOG} shows the mass spectra corresponding to the molecular content associated with a highly ordered pyrolitic graphite (HPOG) sample, which is revealed by one-step LDI. Different values of the desorption laser fluence were used. For \emph{F}$_{des} < 300$\,mJ/cm$^{2}$ almost no signal is observed (except two peaks at the noise level corresponding to C$_{10}$ and C$_{12}$). At \emph{F}$_{des} = 300$\,mJ/cm$^{2}$ a distribution of C clusters is observed, extending to C$_{28}$. Increasing the fluence to \emph{F}$_{des} =1$\,J/cm$^{2}$ leads to higher peaks intensities and a distribution extending to C$_{35}$.
These observations indicate that ablation of carbonaceous material and chemistry within the plume can produce C clusters but not fullerenes. We expect similar results for other carbonaceous materials \citep[e.g. diamond;][]{Sedo2006}. 

We also studied the evolution as a function of laser conditions of the size distribution of fullerenes in the AhS~\#04 sample (see Fig.~\ref{fig:fullerenes_cal}). In this sample, two-step LDI scheme with wavelengths and fluences typical of those used to analyze PAHs and fullerenes in complex hydrocarbon mixtures without significant fragmentation \citep{Faccinetto2011High-sensitivity, Homann1998Fullerenes, Lykke1993Molecular}. 
The fullerene ion signals were summed considering three size ranges covering medium ($42\leq C\# \leq58$), large ($60\leq C\# \leq70$), and very large ($C\#>70$) sizes.
Figure~\ref{fig:fullerenes_cal} shows that the fullerene size distributions remain comparable within a factor of typically 2, independent of the laser conditions.  These calibration measurements support the fact that fullerenes are intrinsic to the samples.
By examining the relative fullerene ion signal between the 7 AhS ureilite samples (Fig.~\ref{fig:fullerenes_cal}), we conclude that the fullerene size distribution is comparable in all samples, with the exception of an obvious lack of very large fullerenes in AhS~\#22 and \#24 in which the detected species only go up to C\#=80.
We cannot conclude whether this difference reflects an intrinsic difference in fullerene sources. Rather, it indicates the difficulty of extracting very large fullerenes from bulk samples and the possible role of sample heterogeneity.

\begin{figure*}[ht]
\epsscale{1.2}
\plotone{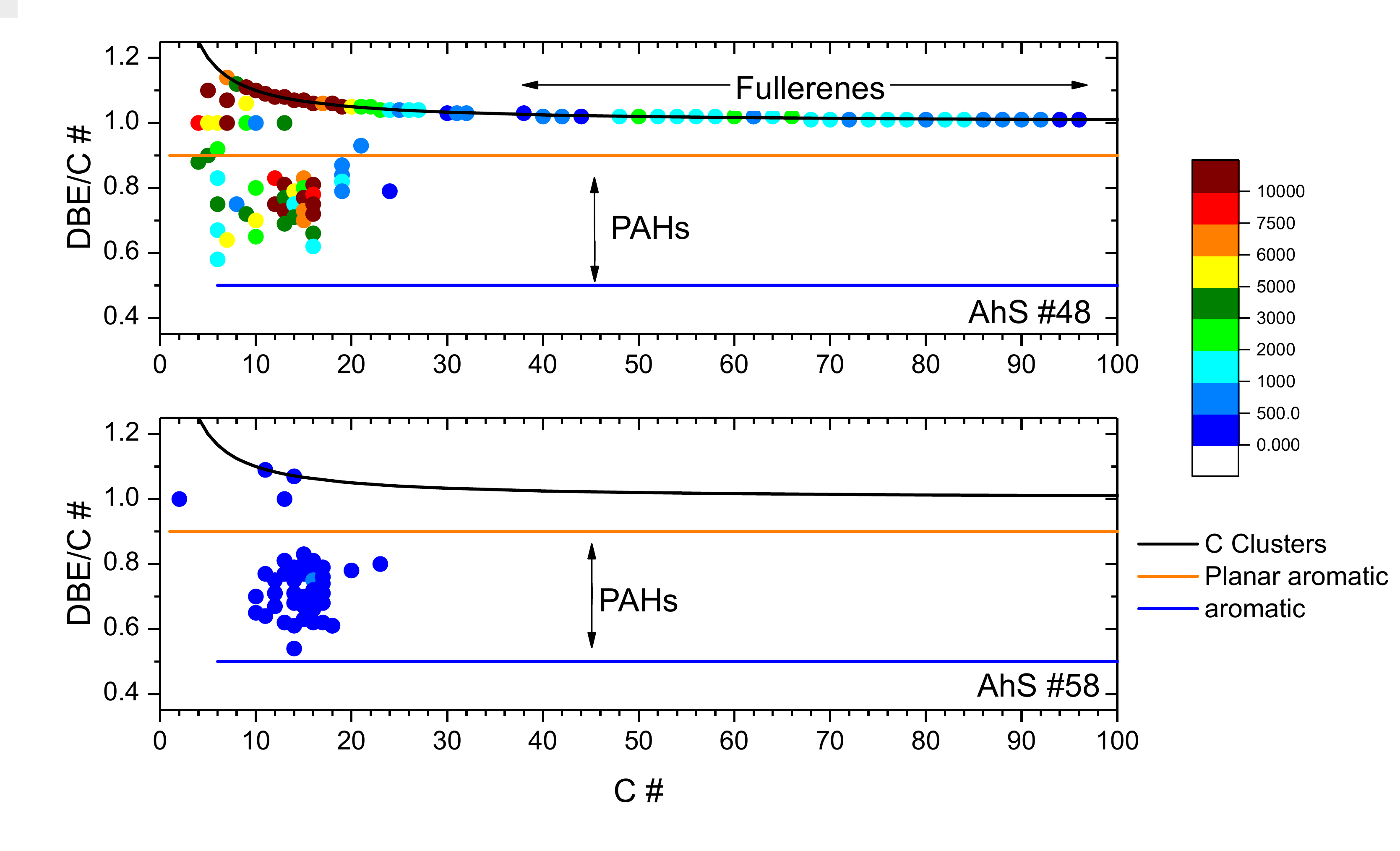}
\caption{Plot of DBE/C\# versus carbon number (C\#) for AhS samples \#48 and \#58. The colored scale corresponds to the peak intensities in the recorded mass spectra. Below DBE/C\# = 0.5 the species are considered aliphatic and above as containing aromatic cycles \citep{Koch2006From}. Aromatics remain planar below DBE/C\#=0.9 \citep{Hsu2011Compositional}. Above this value, the species are classified as HC clusters. The upper black line corresponds to pure C clusters.
\label{fig:DBE}}
\end{figure*}

\begin{figure*}[ht]
\epsscale{1.25}
\plotone{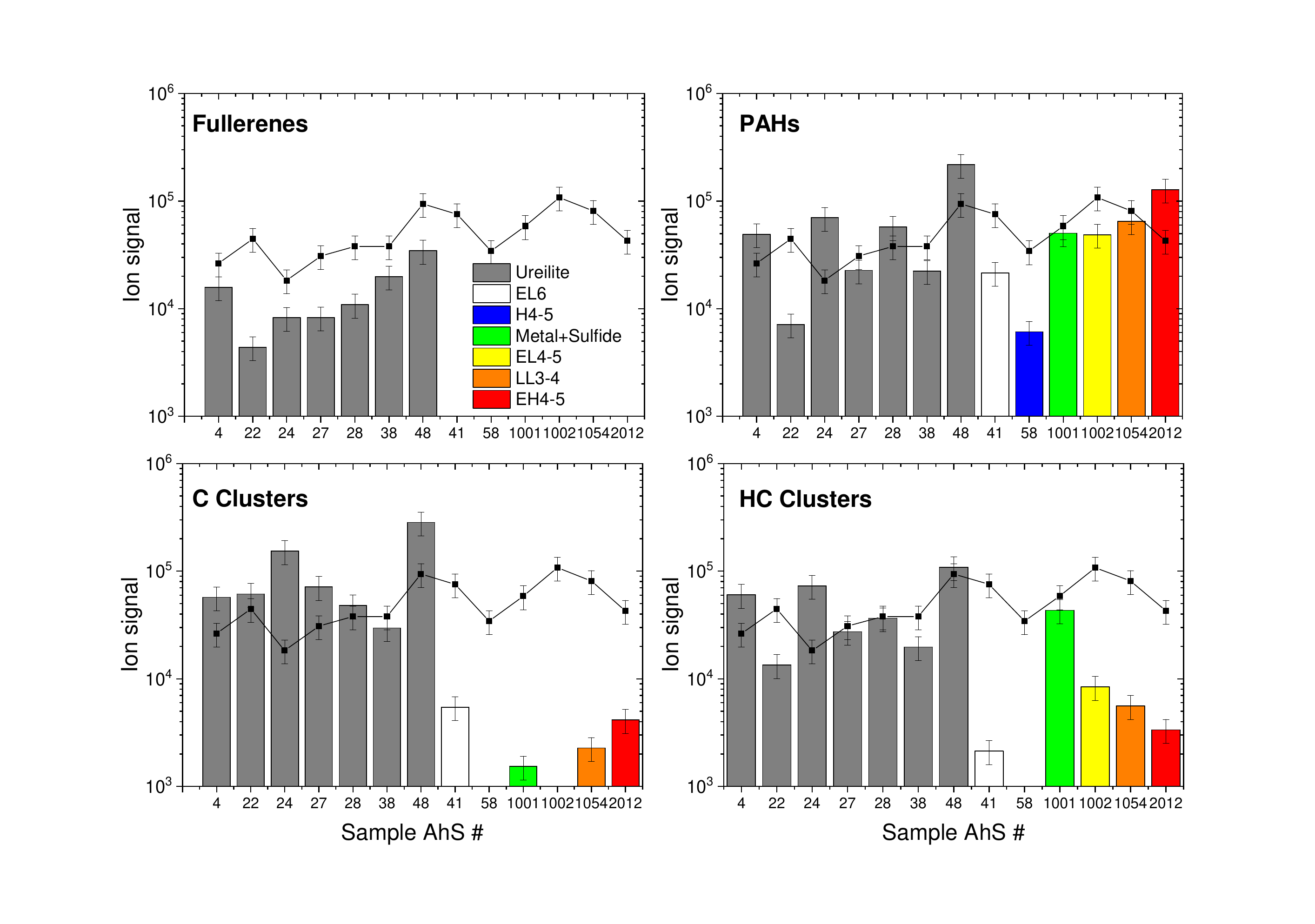}
\caption{Sum of peak intensities in mass spectra recorded after classification into molecular families and for all AhS meteorite clasts studied. Each bar graph gives a picture among the meteorite samples of this sum for each molecular family (fullerenes, C clusters, HC clusters and PAHs) relative to the sum of peak intensities assigned to metals (points). Ureilites are on the left (gray bars) and non-ureilites on the right (other colors). Error bars shown are the relative standard deviation obtained by repeating the measurement at different dates.
\label{fig:family_int}}
\end{figure*}

\begin{figure}[h]
 \centering
        \includegraphics[width=0.6\linewidth]{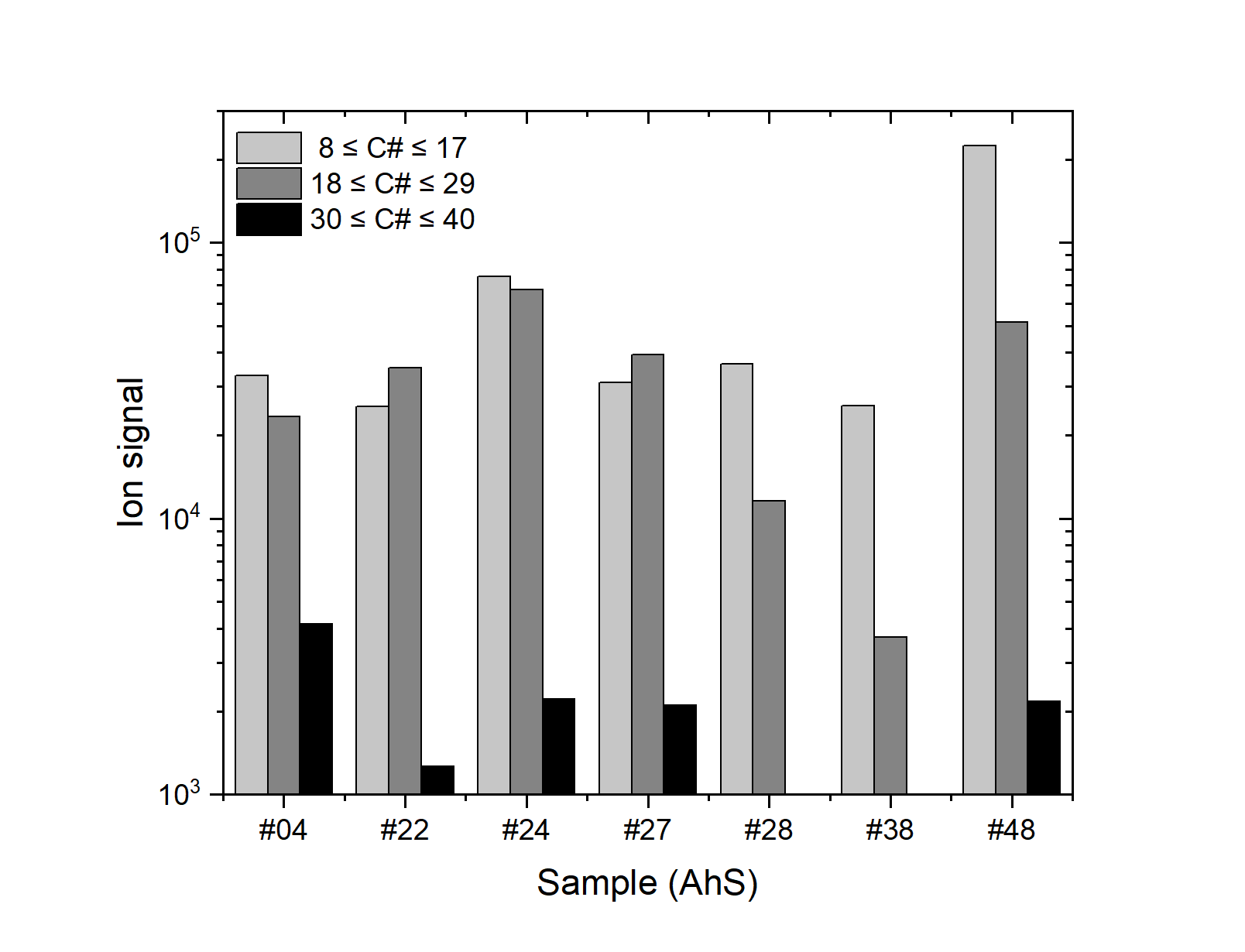}
\caption{Sum of peak intensities for C clusters grouped into two ranges of small ($8\leq C\# \leq 17$) and medium ($18\leq C\# \leq 29$) size as well as  the $30\leq C\# \leq 40$ category that contains both large C clusters and small fullerenes. The values obtained for the 7 ureilite samples are presented.
\label{fig:cc_AhS}}
\end{figure}

\subsection{Classification in molecular families} \label{subsec:families}
Figure~\ref{fig:lowMS} shows that hundreds of peaks are detected in the low m/z mass range [10-500] of the thirteen AhS samples. These peaks include a large diversity of carbonaceous species. We sorted these molecules into families by applying the DBE method (see Section~\ref{subsec:DBE}). Examples of DBE/C\# plots are provided in Figure~\ref{fig:DBE}.
The families observed include aliphatics, HC clusters, PAHs, C clusters, and fullerenes as described in \cite{Sabbah2020Molecular}. 
The aliphatic family is not considered because we could detect only a few species falling into this category, five species in AhS~\#04 and three in AhS~\#1001, all at C\#$\leq 7$.
The sum of peak intensities for the different families are presented in Figure~\ref{fig:family_int} and compared to the sum of peak intensities of the metals. Note that Figure~\ref{fig:family_int} shows that the intensity of metals in non-ureilites is higher compared to ureilites. 
Conversely, the ion signal of carbonaceous species is much weaker in non-ureilites, with the exception of PAHs, which show similar sum of peak intensities in both clast types and HC clusters in AhS~\#1001.
The observation of PAHs in the porous chondritic components of the asteroid was previously explained as being due to the mobilization of organic matter during impacts \citep{Sabbah2010Polycyclic}. The PAHs detected are relatively small with a maximum of C\#=24 and should be much easier to evaporate than the larger fullerenes.

Similar to the fullerene analysis (Fig.~\ref{fig:fullerenes_cal}), we summed the C clusters peak intensities for the two size categories, small ($8\leq C\# \leq 17$) and medium ($18\leq C\# \leq 29$). As shown in Fig.~\ref{fig:cc_AhS}, the ion signal of these species is large and generally exceeds the ion signal of fullerenes by a factor of 10. The contribution to the ion intensity of the family with $30\leq C\# \leq 40$ (which contains both large C clusters and small fullerenes) is small. Adding this contribution to the C clusters family or the fullerenes family, as was done in Fig.~\ref{fig:family_int} should therefore not affect the results.

The observed variations in the sum of peak intensities with cluster sizes (Fig.~\ref{fig:cc_AhS}) may indicate different origins for small and medium-sized clusters. We have seen that, under our experimental conditions, C clusters can be formed by the interaction of the IR laser with a solid carbonaceous material (e.g. HPOG) whereas fullerenes cannot (cf. Section~\ref{subsec:Fullerenes}). This does not exclude a contribution from C clusters trapped in molecular form in the samples, as suggested in our earlier work on laboratory analogs of stardust \citep{Martinez2020Prevalence}. The different possible origins of the carbon cluster ion signal could explain why this signal is not closely correlated with the fullerenes ion signal.
If fullerenes are associated with only one specific carbonaceous phase among several, the fullerene ion signal would depend on the concentration of that phase and its accessibility by the desorption laser to achieve extraction of its molecular content.
Therefore, the apparent absence of fullerenes in non-ureilites could be only a sensitivity problem.

\begin{figure}[h]
 \centering
        \includegraphics[width=0.65\linewidth]{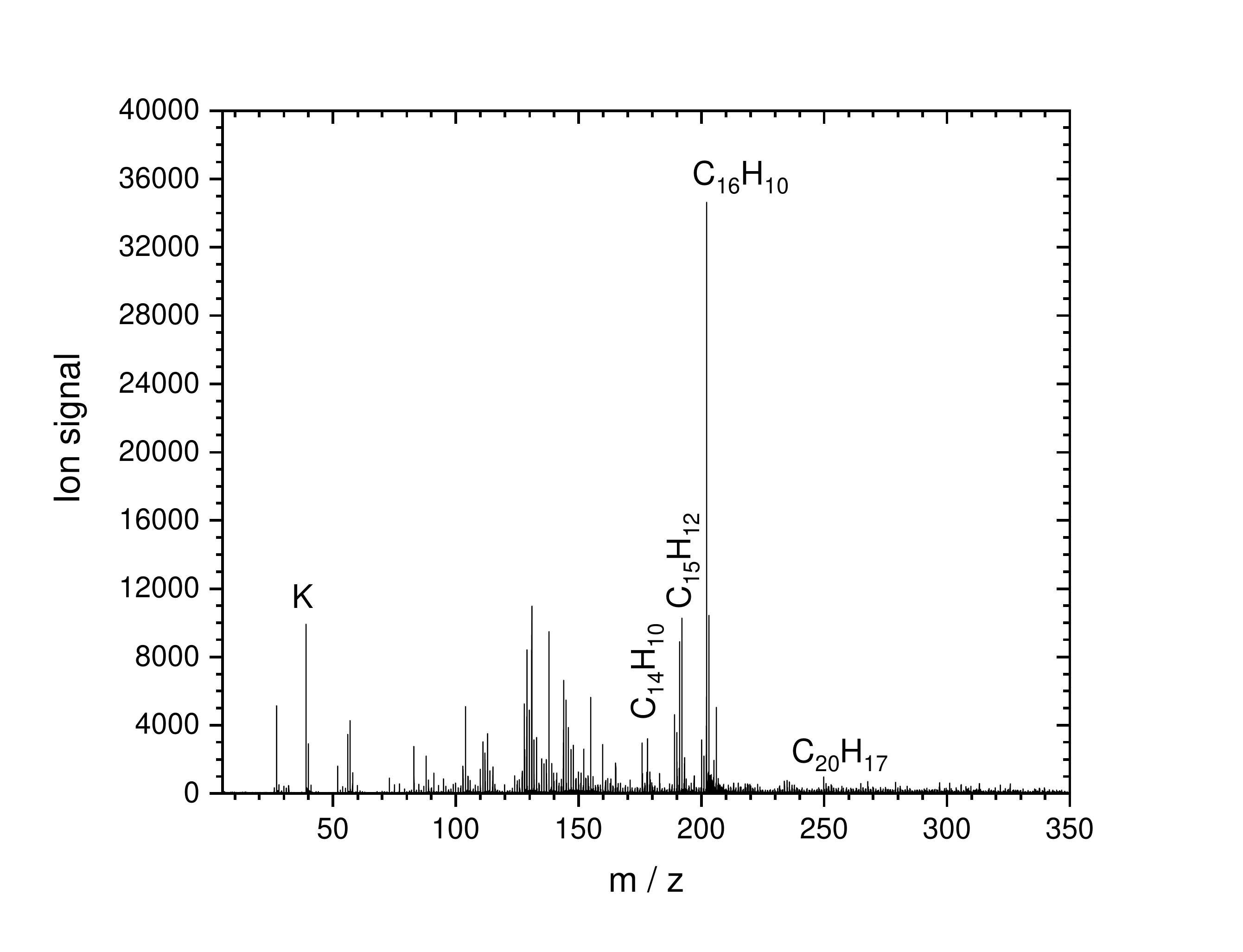}
\caption{Mass spectrum of a bulk sample of the Allende meteorite recorded with the AROMA setup. Only a few m/z peaks are annotated. All identified carbonaceous species are presented in Fig.~\ref{fig:dbe_c_meteorites}. 
\label{fig:Allende}}
\end{figure}
\subsection{The lack of evidence for the presence of fullerenes in the Murchison and Allende meteorites}\label{subsec:fullerenes_allende}

\begin{figure}[htb!]
\epsscale{1.2}
\plotone{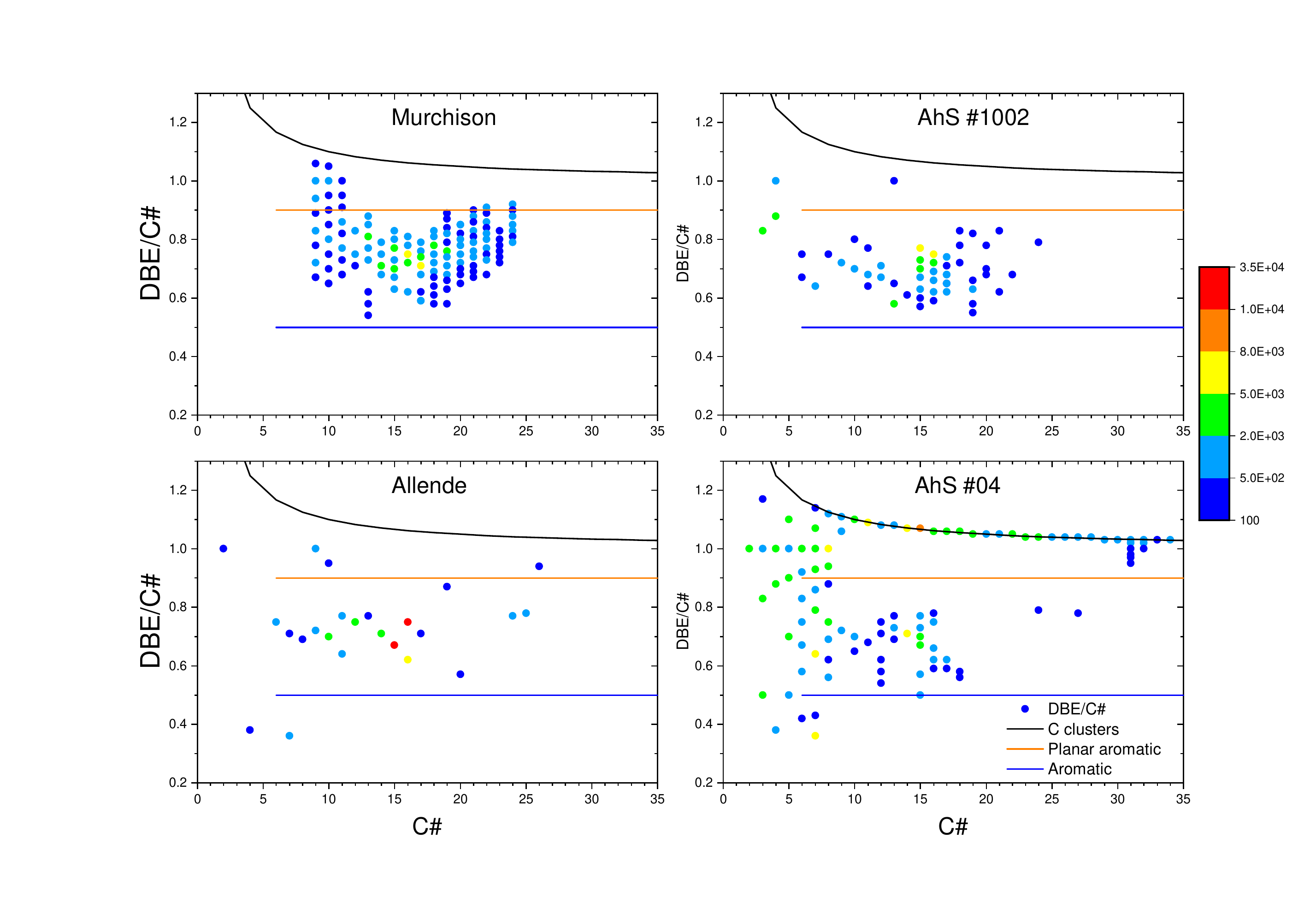}
\caption{Plot of DBE/C\# versus carbon number (C\#) for 4 bulk samples: Murchison, Allende, AhS \#1002 (EL4-5), and \#04 (ureilite). The colored scale refers to peak intensities in the recorded mass spectra. The solid lines show the DBE limits that are used to sort the carbonaceous families (see details in the lower right panel). Note that the mass spectrum of Murchison \citep{Sabbah2017Identification} was not recorded for $m/z<100$.}
\label{fig:dbe_c_meteorites}
\end{figure}

\begin{figure}[htb!]
\plotone{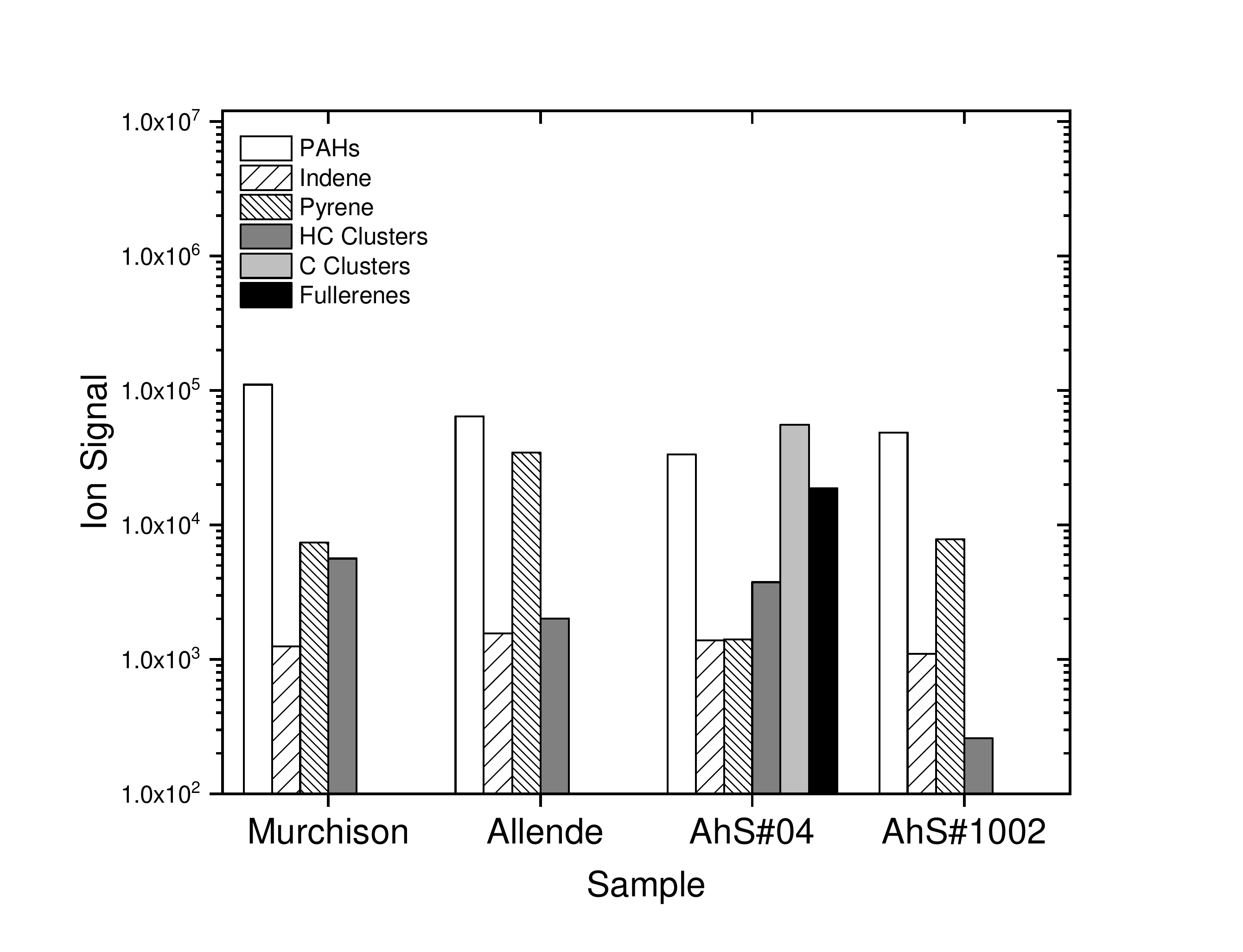}
\caption{Molecular families derived after DBE analysis for Murchison, Allende, AhS~\#1002 (EL4-5), and AhS~\#04 (ureilite). In addition to the sum of the peak intensities for the 4 families (PAHs, HC clusters, C clusters, fullerenes), we report the peak intensities of two individual PAHs: indene (C$_{9}$H$_{8}$) and pyrene (C$_{16}$H$_{10}$).}
\label{fig:family_meteorites}
\end{figure}

Previously reported detections of fullerenes in the Allende \citep{Becker1994} and Murchison meteorites \citep{Becker2000} have been questioned both because of the analytical techniques used \citep{Hammond2008Identifying} and the non-detection of these species by other groups \citep{Buseck2002}. We did not detect fullerenes or C clusters in our previous analysis of a bulk sample of Murchison with the AROMA setup \citep{Sabbah2017Identification}. To complete this result, we analyzed a powder from the interior of the Allende meteorite. The molecular composition of the carbonaceous species (see mass spectrum in Fig.~\ref{fig:Allende}) is dominated by PAHs and a few  m/z peaks attributed to HC clusters (4 peaks) and aliphatics (3 peaks). There is no fullerene signature in the mass spectrum. The PAH distribution peaks around m/z=202.07, which corresponds to pyrene and its isomers. It is found to be similar to that previously detected in several samples from the interior of Allende \citep{zenobi1989}.
Figure~\ref{fig:dbe_c_meteorites} shows the DBE/C\# values in the C\# range of 5 to 35 for the Murchison, Allende, AhS~\#1002, and AhS~\#04 samples. The molecular families of these samples are clearly different, suggesting possible differences in the initial chemical reservoirs or processes undergone by their parent body. Each plot of DBE/C\# pattern is indeed unique. PAHs dominate in three of the four selected samples, but they show some differences. Murchison contains a substantial fraction of methylated PAHs, while Allende contains more pyrene and isomers. In AhS~\#04, the ion signal is comparable for pyrene and indene, the smallest PAH (see Fig.~\ref{fig:dbe_c_meteorites}).  Using the DBE/C\# values, we sorted the carbonaceous molecules into families (for m/z greater than 100) and the results obtained for the four samples of Fig.~\ref{fig:dbe_c_meteorites} are presented in Fig.~\ref{fig:family_meteorites}. The total PAH ion signal in Murchison is the highest among the 4 samples, due to the greater variety of PAHs detected (Fig.~\ref{fig:dbe_c_meteorites}). The fullerene ion signal in AhS~\#04 is lower than that of PAHs in Allende and AhS~\#1002 by a factor of 2-3 and by a factor of 5-6 than that of PAHs in Murchison. It is accompanied by a high signal of C clusters, at the same level as that of PAHs in Allende and AhS~\#1002.

\section{Discussion} \label{sec:discussion}

\subsection{About the concentrations of PAHs, C clusters and fullerenes}\label{subsec:quantification}

Quantifying species abundance as a function of ion signal remains a challenge in LDI techniques \citep{Elsila2004}. The ion signal depends on several parameters, including the desorption and ionization efficiencies of the associated species, the nature of the analyte, the properties of its bonding with the surface, and the nature of the substrate itself. 
Signal dispersion can therefore be attributed to a combined effect of sample heterogeneity and the fact that mainly species from the sample surface are desorbed. This is different from destructive techniques such as ICP-MS using laser ablation in which all material from the laser crater is analyzed \citep{Xiao2021}.
Relating the ion signal to abundances (on the surface) therefore requires calibration tests using several internal standards \citep{Elsila2004} and samples of known concentrations \citep{Koumenis1995Quantitation}.

The inferred mass concentration for PAHs in the Murchison meteorite ranges from 15 to 28 ppm, with an average of 22 ppm \citep{Sephton:2002bc}. Because the Murchison and AhS samples were analyzed with AROMA using the same experimental conditions and similar mass of each meteoritic sample, the sum of peak intensities for PAHs from the two experiments can be directly compared (see Figure~\ref{fig:family_meteorites}). Only two samples show a value slightly higher than the Murchison value, while the others are lower by a factor of about two (six of them) and up to 20 for two of them. Considering the number of factors that can affect the absolute intensities \citep{Elsila2004}, we can conclude that the PAH mass concentrations are lower but comparable in AhS compared to Murchison. 

In a previous study on soot, we demonstrated the ability to identify fullerene species in nascent soot particles produced in a slightly sooting ethylene/air premixed flame \citep{Sabbah2020Molecular}. In addition, we were able to follow the evolution of carbonaceous molecular families, including fullerenes, by analyzing soot collected at different heights above the burner. In this study, we have shown that AROMA can quantitatively trace the thermal processing of large PAHs (C\#$\geq$50) leading to the formation of fullerenes in this hydrocarbon flame and that the detected molecules mainly originate from the surface of soot particles. This suggests that the efficiency of AROMA in detecting both molecular species (i.e. large PAHs and fullerenes) is similar \citep[at least when associated with soot particles;][]{Sabbah2020Molecular}. Assuming this is the case for the AhS samples, the mass concentrations of fullerenes can be deduced by considering that the summed ion signal for fullerenes and PAHs is proportional to their respective concentrations. Therefore, the mass concentration of fullerenes is most probably comparable to that of PAHs in AhS, i.e. a few ppm (after correction of the ratio between the average mass of PAHs (m/z$\sim$212) and that of fullerenes (m/z$\sim$740)). However, this value should be taken with caution because PAHs and fullerenes are distributed in different phases within the AhS meteorite, PAHs being quite widespread while fullerenes are supposed to be associated with a specific solid carbon phase.

\subsection{Scenarios of fullerene formation in AhS}\label{subsec:origin}

In the following, we discuss the pros and cons of possible scenarios that could explain the presence of fullerenes in the AhS meteorite. One possible scenario for the presence of fullerenes in AhS, which is suggested by the unique history of the ureilite parent asteroid \citep{Goodrich2015Origin}, is the introduction by a primitive (CC-like) impactor that was responsible for the catastrophic disruption of the UPB. The timing of the UPB disruption led to the suggestion \citep{Yin2018} that the impactor was an outer SS body, associated with a large-scale migration of outer SS bodies into the inner SS driven by the growth and/or migration of the giant planets during the gaseous disk phase  \citep{Walsh2012Populating}.  Some impactor material may have been added to the ureilitic daughter bodies formed at this time and then redistributed to other clast types by subsequent impact gardening of the regolith. However, we have only observed fullerenes in ureilites and have not been able to detect fullerenes in two well-known CC, Murchison and Allende.
 
An alternative scenario is the possibility that the fullerenes were produced in the ureilites themselves sometime during their history. The UPB was a carbon-rich asteroid that experienced igneous processing involving temperatures up to $\sim$1250-1300$^{\circ}$C \citep{Collinet2020}. During this processing, graphite was formed from the primitive carbon-rich materials that accreted onto the body. However, there is no known mechanism by which the fullerenes could have formed during igneous processing. Subsequently, the UPB experienced a number of shock events, in particular the large impact that resulted in a major disruption of the parent body \citep{Goodrich2015Origin}; such events offer a potential mechanism for the formation of fullerenes. It has been proposed that the diamonds present in most ureilites formed by transformation of graphite during this shock \citep[e.g.][and references therein]{Lipschutz1964, Nabiei2018}. A recent laboratory study \citep{Popov2020} reports a zone of instability of diamond, for a range of temperatures, at pressures of 55 to 115~GPa. This instability leads to the transformation of diamond into fullerene-like onions, which consist of multiple shells whose number decreases with temperature to a minimum value of 2-3 at a temperature of 2400~K or 2127$^{\circ}$C \citep{Popov2020}. This suggests the possibility that fullerenes in ureilites formed almost immediately after the formation of diamonds in an impact environment.  However, it is unlikely that the extreme conditions necessary for the creation of these fullerene-like onions were achieved in ureilites. \cite{Nestola2020} have shown that diamonds in ureilites could have formed by impact shock events at much lower pressure and temperature. Furthermore, onion species formed by this mechanism differ from molecular fullerenes as shown by Raman spectroscopy \citep{Popov2020}. The idea that fullerenes could have formed by hypervelocity impacts in space has been investigated by the analysis of impacts collected on the Long Duration Exposure Facility (LDEF). Fullerenes were detected in a single LDEF impact crater, which motivated experimental simulations \citep{Radicati1994}. Although these experiments were limited to impacts at a velocity of 6\,km\,s$^{-1}$, the authors concluded that the fullerenes observed in the LDEF crater are unlikely to have formed during a hypervelocity impact. Since this crater also contained chondritic elements, these authors suggested that the fullerenes were instead intrinsic to the impacting micrometeoroid.

Another scenario would be that these fullerenes formed much earlier and were inherited from the interstellar medium at the time of SS formation.  The abundance of carbon in the form of C$_{60}$ in the diffuse interstellar medium is estimated to be on the order of a few 10$^{-4}$ to a few 10$^{-3}$ \citep{Berne2017}. This value is substantially lower than the abundance value of 10$^{-1}$ for interstellar PAHs, which are the carriers of the aromatic infrared bands. Interstellar PAHs are expected to be large, typically containing more than 50 C due to processing by UV photons \citep[e.g.][]{Montillaud2013}. Such large PAHs are not detected in our study. We therefore expect the concentration of interstellar PAHs and/or fullerenes to be very low in AhS (as in Murchison and Allende). Another possibility is that these species are included in a phase from which they cannot be extracted by the AROMA desorption laser. The small PAHs (less than 30 carbon atoms in size) detected in AhS (as well as in Murchison and Allende) are more likely products of active cold chemistry in the dense molecular clouds from which SS formed. This suggestion of \cite{Woods2007} was supported by the recent detection of small PAHs in the cold prestellar core TMC-1, in particular indene C$_9$H$_8$ \citep{Cernicharo2021,Burkhardt2021}, as well as the availability of chemical pathways to form these small aromatic species by neutral neutral reactions at low temperature \citep{Cernicharo2021, Cernicharo2021b}.

Finally, a local interstellar heritage has been proposed by several authors to account for anomalies in short-lived radionuclides (SLRs) such as $^{26}$Al \citep[e.g.][]{Podosek2005Overview, Huss2009}. More precisely, the idea arose that the SS prenatal cloud was seeded by small dust grains produced at the end of the life of nearby very massive stars \citep{Arnould2006production, Dauphas2010Neutron-rich,Gaidos200926Al, Gounelle2012, Tatischeff2010}. It is interesting to note that a production of fullerenes at the end of the life of very massive stars, at the presupernova stage of WR stars or at the supernova stage has been suggested by modelers \citep{Cherchneff2000Dust,Clayton2001, Clayton2018} as discussed in Sect.~\ref{sec:intro}. The model by \citeauthor{Clayton2018} indicates that C-rich regions are strongly enriched in $^{12}$C, although processes leading to mixing with $^{13}$C may occur.
This would help to understand the fact that, although some of the presolar grains of SN origin are strongly enriched in $^{12}$C, a significant fraction of them have an isotopic ratio close to the solar value \citep{Lodders2005Presolar,Croat2003}. We derived a value of 0.68$\pm$0.04 for the ratio of $^{13}$C$^{12}$C$_{59}$/$^{12}$C$_{60}$ for the 7 AhS samples, to be compared with the value of 0.62$\pm$0.04 derived from molecular analysis of terrestrial samples. The two ratios are therefore similar. A systematic study would be needed to refine these values and extend them to the entire fullerene population. However, we emphasize that the inferred $^{12}$C/$^{13}$C ratio may be underestimated due to a possible contribution from fullerane species \citep{Becker1997}.
At this stage, a scenario in which the fullerenes present in AhS originate from massive stars can therefore not be demonstrated on the basis of these isotopic ratios or other arguments. But neither can it be contradicted. This scenario has obvious advantages from an astrochemical perspective. It could rationalize why C$_{60}$ is observed in some astronomical environments and not in others. Its detection is mainly in evolved stars, but in a limited number of them \citep[a few \% of planetary nebulae;][]{Otsuka2014} including H-rich but not H-poor circumstellar environments \citep{GarciaHernandez2011}. C$_{60}$ has also been detected in the environment of a number of massive young stellar objects \citep{Roberts2012Detection, Sellgren2010}. This diversity is difficult to rationalize and may indicate star-forming regions associated with the shell of a WR bubble \citep{Dwarkadas2017Triggered}.

\section{Conclusion}\label{sec:conclusion}
This work was motivated by the previous detection of PAHs and molecules containing only carbon in the AhS meteorite. Using the L2MS technique on bulk samples, we showed that all AhS samples of ureilite type exhibit a large size distribution of fullerenes. We performed different calibration experiments to demonstrate that these fullerenes are indeed intrinsic to the samples and are probably associated with a carbon phase that also produces C clusters under exposure to the desorption laser. The inferred concentration of fullerenes is most likely on the order of a few ppm. Using the same experimental conditions, we were unable to detect fullerenes in the Murchison and Allende carbonaceous chondrites. Further investigation of the carbonaceous phase in which the fullerenes are found is essential to improve the sensitivity of the detection and constrain the formation scenarios of these fullerenes.

Since the UPB was disrupted by a giant impact and the re-accumulated debris then experienced further smaller impacts, it is possible that the fullerenes in AhS originated from synthesis driven by impact shocks. However, there is no experimental data demonstrating that this can be achieved. We put forward the alternative possibility that the fullerenes detected in this work could be of interstellar heritage. The remarkable thermal stability of fullerenes, both in the gas phase and in solid particles, for temperatures up to at least 2500~K \citep{Sommer1996,Biennier2017} may have allowed these primitive species to survive the igneous processing conditions of the parent body as well as the impact shock events experienced by the ureilites. These fullerenes may have formed in late phases of a massive (WR) star that ended its life near the parent molecular cloud of the SS.
 
The ideas discussed here would deserve further investigation. In particular, efforts to search for fullerenes, especially in the most primitive meteorites, should be renewed.
Studying carbon-enriched phases to look for fullerenes could still reveal these species in primitive CCs and confirm previous work by Becker et al. who used such a strategy in their measurements. The detection of fullerenes may have implications not only for the SS formation but also for our understanding of the origin of fullerenes in astrophysical environments, a question widely debated in the literature and one that will certainly progress with the upcoming James Webb Space Telescope observations.

\section*{acknowledgments}\label{sec:acknowledgments}
The authors thank the colleagues who kindly provided some of the samples used in this study, most notably Jes\'us Mart\'inez Fr\'ias and Jos\'e Cernicharo for the Allende sample, and Jos\'e \'Angel Mart\'in-Gago for HPOG. Funding: The research leading to these results has received funding from the European Research Council under the European Union’s Seventh Framework Programme (FP/2007-2013) ERC-2013-SyG, Grant agreement N$^{o}$610256 NANOCOSMOS. MC also acknowledges support from La Region Occitanie, Grant N° 15066466. CAG and PJ are supported by NASA Emerging Worlds program grant 80NSSC19K0513 and its pilot study 00A062101. 

\bibliography{AhSfullerenes}{}
\bibliographystyle{aasjournal}



\end{document}